\documentclass[]{article}
\pdfoutput=1

\usepackage[top=70pt,bottom=70pt,left=90pt,right=90pt]{geometry}
\usepackage{amssymb}
\usepackage[usenames,dvipsnames,svgnames,table]{xcolor}
\usepackage{url}
\usepackage{blkarray}
\usepackage{dsfont}
\usepackage{cite}
\usepackage[hypertexnames=false,hyperfootnotes=false,colorlinks=true,linkcolor=blue,%
citecolor=ForestGreen,filecolor=magenta,urlcolor=cyan,linktocpage=true]{hyperref} 

\newcommand{\norm}[1]{\left\lVert#1\right\rVert}

\usepackage{mathtools}
\DeclarePairedDelimiter\bra{\langle}{\rvert}
\DeclarePairedDelimiter\ket{\lvert}{\rangle}
\DeclarePairedDelimiterX\braket[2]{\langle}{\rangle}{#1 \delimsize\vert #2}

\usepackage{enumitem}
\newlist{checklist}{itemize}{2}
\setlist[checklist]{label=$\square$}
\usepackage{pifont}
\newcommand{\cmark}{\ding{51}}%
\newcommand{\done}{\rlap{$\square$}{\raisebox{2pt}{\large\hspace{1pt}\cmark}}%
\hspace{-2.5pt}}

\newcommand{\1}{ {\color{Bittersweet} 1} }

\newcommand{\A}{ \Omega }
\newcommand{\iswap}{\texttt{iSWAP}}

\begin{document}

\graphicspath{{./img/}}

\title{Quantum gates by resonantly driving many-body eigenstates, with a focus on Polychronakos' model}
\author{Koen {Groenland}$^{1,2,3}$ and Kareljan {Schoutens}$^{1,2}$}
\date{18 January 2019} 
\maketitle

{ \footnotesize{ \noindent
{$^1$ QuSoft, Science Park 123, 1098 XG Amsterdam, the Netherlands} \\
{$^2$ Institute of Physics, University of Amsterdam, Science Park 904, 1098 XH Amsterdam, the Netherlands} \\
{ $^3$ Centrum Wiskunde en Informatica (CWI), Science Park 123, 1098 XG Amsterdam, the Netherlands} 
} \\
Contact: koen.groenland@gmail.com
 }

\begin{abstract}
{
Accurate, nontrivial quantum operations on many qubits are experimentally challenging. As opposed to the standard approach of compiling larger unitaries into sequences of 2-qubit gates, we propose a protocol on Hamiltonian control fields which implements highly selective multi-qubit gates in a strongly-coupled many-body quantum system. We exploit the selectiveness of resonant driving to exchange only $2$ out of $2^N$ eigenstates of some background Hamiltonian, and discuss a basis transformation, the eigengate, that makes this operation relevant to the computational basis. The latter has a second use as a Hahn echo which undoes the dynamical phases due to the background Hamiltonian. We find that the error of such protocols scales favourably with the gate time as $t^{-2}$, but the protocol becomes inefficient with a growing number of qubits $N$. 
The framework is numerically tested in the context of a spin chain model first described by Polychronakos, for which we show that an earlier solution method naturally gives rise to an eigengate.  Our techniques could be of independent interest for the theory of driven many-body systems. 
}
\end{abstract}

\newpage
\tableofcontents

\newpage

\section{Introduction}
Resonant driving techniques are well-known in atomic physics, where they are used to populate specific orbitals \cite{Griffiths2005, Wollenhaupt2016}, and in experimental quantum information processing, where they are exploited to form quantum gates on one or two qubits \cite{Cory1997,Garcia-Ripoll2003,Gambetta2017,Zajac2018}. When a pair of eigenstates with a unique energy gap is resonantly driven for an appropriate amount of time, the unitary time-evolution operator, in the asymptotic limit of increasingly weak driving, approaches the form
\begin{align}
\iswap_{t_1, t_2} = \begin{pmatrix}
\ddots & 	& 	&		& \\
	& 0 & \hdots & \1 e^{i \alpha}	&	\\
	& \vdots	& \ddots & \vdots &\\
	& \1 e^{i \beta} &  \hdots & 0 & \\
	&	& &  	& \ddots
\end{pmatrix}.
\label{eqn:udrive}
\end{align}
Here, all diagonal entries are 1, except in the subspace spanned by the resonant states, which we denote by $t_1$ and $t_2$.

We observe that this is very similar to frequently encountered many-qubit gates in quantum information processing, such as the universal Toffoli gate (a bitflip $\sigma^x$ on a target qubit if and only if two control qubits are in the state $\ket{1}$) and Fredkin gate (swapping the states of two target qubits, if and only if a control qubit is $\ket{1}$). Both gates also have the property that all diagonal entries are $1$, except in a two-dimensional subspace:
\begin{align*}
\iswap_{110, 111} = \text{Toffoli} = 
\begin{blockarray}{ccccccccc}
  & 000 & 001 & 010 & 011 & 100 & 101 & 110 & 111 \\
\begin{block}{c(cccccccc)}
 000 & \1 & 0 & 0 & 0 & 0 & 0 & 0 & 0 \\
 001 & 0 & \1 & 0 & 0 & 0 & 0 & 0 & 0 \\
 010 & 0 & 0 & \1 & 0 & 0 & 0 & 0 & 0 \\
 011 & 0 & 0 & 0 & \1 & 0 & 0 & 0 & 0 \\
 100 & 0 & 0 & 0 & 0 & \1 & 0 & 0 & 0 \\
 101 & 0 & 0 & 0 & 0 & 0 & \1 & 0 & 0 \\
 110 & 0 & 0 & 0 & 0 & 0 & 0 & 0 & \1 \\
 111 & 0 & 0 & 0 & 0 & 0 & 0 & \1 & 0 \\
\end{block}
\end{blockarray} \\
\iswap_{110, 101} = \text{Fredkin} = 
\begin{blockarray}{ccccccccc}
  & 000 & 001 & 010 & 011 & 100 & 101 & 110 & 111 \\
\begin{block}{c(cccccccc)}
 000 & \1 & 0 & 0 & 0 & 0 & 0 & 0 & 0 \\
 001 & 0 & \1 & 0 & 0 & 0 & 0 & 0 & 0 \\
 010 & 0 & 0 & \1 & 0 & 0 & 0 & 0 & 0 \\
 011 & 0 & 0 & 0 & \1 & 0 & 0 & 0 & 0 \\
 100 & 0 & 0 & 0 & 0 & \1 & 0 & 0 & 0 \\
 101 & 0 & 0 & 0 & 0 & 0 & 0 & \1 & 0 \\
 110 & 0 & 0 & 0 & 0 & 0 & \1 & 0 & 0 \\
 111 & 0 & 0 & 0 & 0 & 0 & 0 & 0 & \1 \\
\end{block}
\end{blockarray}
\end{align*}
An obvious generalization of both Fredkin and Toffoli would include more control qubits, which allow the transition if and only if these are $\ket{1}$. A quantum circuit equivalent to such generalizations using $N$ control qubits has a depth of $O(N)$. 

Note that it is generally highly nontrivial to form gates of the type $\iswap_{t_1, t_2}$ using a local Hamiltonian. They could in principle be generated by a time-independent Hamiltonian of the form $H = \ket{t_1}\bra{t_2} + h.c.$, but such interactions, which act only on many-particle states $t_1$ and $t_2$ but not any others, are typically highly nonlocal, and hence are never encountered in nature \cite{Preskill2013}. When restricting to realistic $2$-local Hamiltonians, in which each term is allowed to act non-trivially on at most $2$ qubits, time-dependent control fields are required. Our goal is to cleverly engineer $2$-local Hamiltonians whose time evolution swaps just 2 out of $2^N$ states and leave all other states put, without resorting to discrete gate decompositions. To do so, we employ systems of the form
\begin{align*}
H(t) = H_\text{bg} + \A' \cos(\omega t + \phi) H_\text{\text{drive}},  
\end{align*}
where $H_\text{bg}$ is a background Hamiltonian whose eigenstates are known, and $H_\text{\text{drive}}$ is some local driving field which incites a transition between two eigenstates of $H_\text{bg}$, which we will call $\ket{t_1}_{H_\text{bg}}$ and $\ket{t_2}_{H_\text{bg}}$. If these eigenstates have a unique energy gap, the resulting operation $U_\text{drive}$ can be made to look as in Eq. \ref{eqn:udrive}. 

Note that the resulting $U_\text{drive}$ has this special form in the \emph{eigenbasis} of $H_\text{bg}$. For further quantum information processing, we propose an operation which maps each eigenstate to a unique computational basis vector, which we call the \emph{eigengate} $U_\text{eg}$. The complete protocol is then described by 
\begin{align*}
\iswap_{t_1,t_2} \approx U_\text{eg} U_\text{drive} U_\text{eg}^\dagger.
\end{align*}
An implementation of this protocol would require the following ingredients:
  \begin{checklist}
  \item A constantly applied background Hamiltonian $H_{\text{bg}}$ which has a unique energy gap $\omega$ between two eigenstates $\ket{t_1}_{H_\text{bg}}$ and $\ket{t_2}_{H_\text{bg}}$. 
  \item A driving field $H_{\text{drive}}$ which couples the states $\ket{t_1}_{H_\text{bg}}$ and $\ket{t_2}_{H_\text{bg}}$, whose amplitude can be made oscillatory at the right frequency $\omega$.
  \item An operation which maps (any) two computational basis states, call them $\ket{t_1}$ and $\ket{t_2}$, to energy eigenstates $\ket{t_1}_{H_\text{bg}}$ and $\ket{t_2}_{H_\text{bg}}$ respectively. We also need the inverse of this operation. 
    \item An efficient method to keep track of the dynamical phases due to $H_{\text{bg}}$. 
  \end{checklist}
Throughout this paper, we will elaborate on the above four checkboxes, and use the example of a spin chain model first described by Polychronakos \cite{Polychronakos1993} for which we show that all requirements can be fulfilled. 

The resulting operation could find applications in noisy intermediate-scale quantum computers \cite{Preskill2018}, where decoherence prohibits long gate sequences, but evolution by engineered Hamiltonians might be natively available. We find that there is a trade-off between gate time and fidelity, where the error $\mathcal{E} \propto t_d^{-2}$ scales as the inverse-square of the driving duration $t_d$. Moreover, as the number of involved qubits increases, the performance of our gate quickly degrades, meaning that conventional gate decompositions are preferred in the many-qubit limit. However, we find that for a modest number of qubits, our gate can be competitive with conventional methods.

\subsection{Related work}
We previously described a very similar resonantly driven gate in Ref. \cite{Groenland2018}, which was based on the so-called Krawtchouk spin chain. In the present work, we generalize many aspects of this first result, and show how the same line of reasoning applies to a very different system featuring long-range rather than just nearest-neighbor interactions. 

In 2010, two independent groups \cite{Burgarth2010, Kay2010} described a result that similarly exploits resonant driving in a many-body system to construct quantum gates. They considered an interacting spin chain of which only the first one or two sites can be controlled, and prove that universal operations over all states are possible. The scope of these papers, achieving single- and two-qubit gates with limited control, is very different from our goal, which is the creation of an unconventional multi-qubit operation on a system with much more generous controllability. Moreover, both earlier papers focus mostly on proof of existence, and do not mention concrete examples of systems which would allow their protocol. In this work, we present an experimentally feasible example which we simulate numerically. 

The most obvious competitor of our protocol is conventional compiling of any quantum operation into a universal set of single- and two-qubit gates. Extensive research efforts have greatly optmized compiling methods, and in the asymptotics of many qubits, compiling approach becomes increasingly favorable compared to our proposal. For a recent overview, see Ref. \cite{Chong2017}. We present our work not as an alternative to compiling, but rather as a creative twist to the fields of condensed matter and quantum control, which might find applications on highly specialized systems. We also present our methods, such as the eigengate presented in Sec. \ref{sec:eg}, as tools that may find applications elsewhere.

\subsection{Document structure}
In section \ref{sec:eg}, we discuss the creation of eigenstates of $H_\text{bg}$ in a general setting, after which we address resonant driving and the errors introduced due to spectator states section \ref{sec:resdrive}. In section \ref{sec:pc} we show how our protocol can be implemented in Polychronakos' model, and in section \ref{sec:numerics} we study the resulting gate numerically. Finally, in section \ref{sec:discussion}, we return to the requirement checklist and discuss to what extend the requirements were satisfied, followed by a conclusion in section \ref{sec:conclusion}.

\section{Mapping the computational basis to eigenstates}
\label{sec:eg}

\subsection{Eigengates from quenches}

\begin{figure}
\centering
\def\svgwidth{.25\linewidth}
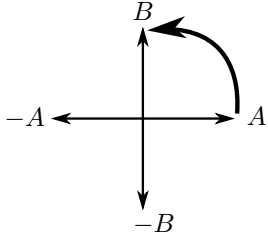
\caption{An eigengate generated by $H$ rotates operator $A$ between $B$, $-A$, $-B$ and back into $A$. Note that consecutive application of two eigengates inverts the spectrum of $A$ or $B$. }
\label{fig:eg_quench_rotation}
\end{figure}

Let $A$ and $B$ be Hermitean operators (or Hamiltonians). We call $U_\text{\text{eg}}$ an \emph{eigengate} between $A$ and $B$ if it maps every eigenstate of $A$ to an eigenstate of $B$. Such eigengates can be implemented by \textit{quenching} (suddenly applying) a third Hamiltonian $H$ which satisfies
\begin{align}
e^{-i H \frac{\pi}{2}}  A e^{i H \frac{\pi}{2}} = B.
\label{eqn:eg_heisenberg}
\end{align}
It is not clear which tuples $(A,B,H)$ satisfy Eq. \ref{eqn:eg_heisenberg} in general, but we claim the following sufficient condition:
\begin{align}
[H,A] &=  i B \quad \text{ and } \quad [H,B] =  -i A.
\label{eqn:alt_comm}
\end{align}
Note that one can freely rescale $A$, $B$ and $H$ (together with the rotation angle $\pi/2$), as this does not change the eigenstates, hence pertaining the same eigengate. To prove our claim, we first recall the Hadamard Lemma, 
\begin{align}
e^{-i H t} A e^{i H t} &= \sum_{n=0}^{\infty} \frac{(-it)^n}{n!} (\text{ad}_H)^n A     \label{eqn:hadlemma}  \\
\text{ad}_H A &:= [H, A]  \nonumber
\end{align}
and plug Eq. \ref{eqn:alt_comm} into Eq. \ref{eqn:hadlemma}: 
\begin{align*}
e^{-i H t}  A e^{i H t} &= A + (-it) (iB) + \frac{(-it)^2}{2!} A + \frac{(-it)^3}{3!} (iB) + \frac{(-it)^4}{4!} A + ...  \\
&= \sum_{k = 0}^\infty \frac{(-it)^{2k}}{2k!} A 
+  \sum_{k = 0}^\infty \frac{(-it)^{2k+1}}{(2k+1)!}i B \\
&= \cos( t ) A + \sin( t ) B .
\end{align*}
Hence, the operator $A$ evolves to $B$ in the Heisenberg-picture if we apply $H$ for a time $t= \frac{\pi}{2}$. We graphically depict such rotations in Fig. \ref{fig:eg_quench_rotation}. Let the subscripts of kets denote the basis in which the vector is described. Then, for every eigenvector $\ket{j}_A$ of $A$, we may define 
\begin{align*}
\ket{j}_B = U_\text{eg} \ket{j}_A
\end{align*}
where $U_\text{eg} = \exp( -i H \frac{\pi}{2} )$ is an eigengate from $A$ to $B$. Unitaries are isospectral mappings, preserving the spectrum of Hermitean operators, hence for all $j$, if $A \ket{j}_A = \lambda_j \ket{j}_A$, then $B \ket{j}_B = \lambda_j \ket{j}_B$. Examples of valid $(A,B,H)$ tuples include $(\sigma^x, \sigma^y, \sigma^z)$ where $\sigma^\alpha$ denote the Pauli matrices, and $(L_0^\alpha, L_1^\alpha, H_\text{P} )$ as we will discuss in Sec. \ref{sec:pc} on Polychronakos' model. 

There is some leftover symmetry $H \rightarrow H+G$ for any $G$ such that $[G,A] = [G,B]=0$. In other words, within each subspace spanned by eigenvectors of $A$ with the same eigenvalue, the operator $U_\text{eg}$ may cause an arbitrary unitary rotation which we cannot track using the method presented here. 

\subsection{Eigengates from adiabatic evolution}
Another method to turn eigenstates of $A$ into eigenstates of $B$ is by adiabatic evolution. For $t \in [0, \pi/2]$, we consider the adiabatic Hamiltonian 
\begin{align}
H_\text{adiabatic} = \cos( t ) A + \sin( t) B.
\label{eqn:adiabatic}
\end{align}
Because the LHS of Eq. \ref{eqn:hadlemma} shows an isospectral transformation of $A$, the eigenvalues of Eq. \ref{eqn:adiabatic} remain the same at all times. Assuming the relevant units of energy are large compared to the units of time, the total time-evolution converges to the form 
\begin{align*}
U_\text{eg,adiabatic} &= \mathcal{T} \exp \left(-i \int_0^{\frac{\pi}{2}} H_\text{adiabatic}(t) dt \right) \\
&= \underbrace{ \exp \left( -i G \frac{\pi}{2} \right)  }_\text{Acts only within degenerate subspaces}  \underbrace{ \exp \left( -i H \frac{\pi}{2} \right)  }_{\text{Eigengate} }  \underbrace{ \exp \left( -i A \frac{\pi}{2} \right)  }_{\text{Dynamical phase} }
\end{align*}
where the form of $G$ in the last equation is unknown, but limited to act nontrivially only within each subspace of fixed eigenvalue. On individual eigenstates of $A$, the adiabatic evolution operator acts as
\begin{align*}
U_\text{eg,adiabatic} \ket{j}_A = \exp \left(-i G \frac{\pi}{2} \right) \exp \left( -i \lambda_j \frac{\pi}{2} \right) \ket{j}_B.
\end{align*}

\subsection{Eigengates for resonantly driven transitions}
Our goal is to arrive at a quantum gate that exchanges exactly two states in the \emph{computational basis} by driving a unique transition in some background Hamiltonian's \emph{eigenbasis}. This is where eigengates come in. 

We require $A$ to be \textit{any} Hamiltonian which is diagonal in the computational basis, while $B$ is the background Hamiltonian in which the driving takes place. In that case, the eigengate $U_\text{eg}$ between $A$ and $B$ maps computational states to eigenstates. Using states in the eigenbasis of $B$, we may selectively exchange eigenstates using resonant driving. Finally, an inverse eigengate (or equivalently, a $\frac{3 \pi}{2}$ rotation by $H$) then maps back to the computational basis, giving the desired result. 

Although sufficient and highly convenient, the eigengate is not \emph{necessary} for this protocol: any unitary map that sends two computational basis states to the transitioning states would be sufficient. 

As an interesting aside, the eigengate's feature to invert the energy spectrum when employed twice, has applications in perfect state transfer \cite{Christandl2004,Bose2007}. In particular, whenever $A$ is a sum of single-qubit terms such that an excitation at qubit $i$ has energy opposite to an excitation at qubit $j$, then $(U_\text{eg})^2$ exchanges the state of qubits $i$ and $j$. This assumes conservation of the number of excitations, and the uniqueness of the relevant energies.

\section{Resonant driving}
\label{sec:resdrive}
This section studies the effect of resonant driving on many-body systems. We now consider a Hamiltonian of the form 
\begin{align}
H(t) = H_\text{bg} + \A' \cos(\omega t + \phi) H_\text{\text{drive}}  
\label{eqn:generaldriving}
\end{align}
where $\A' \ll 1$ and $\norm{H_\text{bg}} \sim \norm{H_\text{\text{drive}}}$, justifying a perturbative treatment of the second term. To leading order in $\A'$, we consider all pairs of eigenstates of $H_\text{bg}$, where each pair approximately forms an independently interacting two-level system. We will argue that, for sufficiently small $\A'$, no transitions between eigenstates of $H_\text{bg}$ occur unless $\omega$ is close to any energy gap between any pair of eigenstates. All states that are not involved in such transitions will be named spectator states, which merely pick up a dynamical phase.

\subsection{A toy example: driving a two-level system}
Consider a toy model in which any pair of eigenstates is described by the qubit system
\begin{align}
H(t) &= \frac{\Delta E}{2} \sigma^z + \A ( \cos(\omega t + \phi) \sigma^x + \sin(\omega t + \phi) \sigma^y )   \label{eqn:tlsdriving} \\
&= \begin{pmatrix}
\frac{\Delta E}{2} & \A e^{-i \omega t -i \phi } \\
\A e^{ +i \omega t + i \phi} & - \frac{\Delta E}{2}
\end{pmatrix}, \nonumber
\end{align}
where $\vec{\sigma} = ( \sigma^x, \sigma^y, \sigma^z )^T$ denotes the vector of Pauli matrices. We can make this Hamiltonian time-independent by moving to the rotating (or interaction) frame, using the basis-transformation 
\begin{align*}
U_\text{rf} = \exp(i \omega \sigma^z t /2 )  = 
\begin{pmatrix}
e^{i \frac{\omega t}{2} } & 0 \\
0 & e^{-i \frac{\omega t}{2} }
\end{pmatrix},
\end{align*}
such that the effective Hamiltonian in this frame becomes 
\begin{align*}
H_\text{rf} &= U_\text{rf} H(t) U_\text{rf}^\dagger + i \dot{U_\text{rf}} U_\text{rf}^\dagger \\
	&= (\Delta E - \omega )\frac{\sigma^z}{2} + \A (e^{i \phi} \sigma^+ + e^{- i \phi} \sigma^- ) \\
	&= \begin{pmatrix}
\delta / 2 & \A e^{-i \phi} \\
\A e^{+i \phi} & -\delta / 2
\end{pmatrix}.
\end{align*}
In the last step, we defined the detuning $\delta = \Delta E - \omega$. Evolution by $H_\text{rf}$ dictates Rabi flopping:
\begin{align}
U_\text{drive}^{\text{(rf)}}( \phi, t ) = e^{-i H_\text{rf} t } = \cos( n t ) \mathds{1} - i \sin( n t ) \left( \frac{ \vec{n} }{n} \right) \cdot \vec{\sigma}  \label{eqn:Udrive} \\
\text{with } \vec{n} = \begin{pmatrix} \cos(\phi) \A \\ \sin(\phi) \A \\ \delta / 2 \end{pmatrix}, 
\quad n = \sqrt{\A^2 + \frac{\delta^2}{4}}.  \nonumber
\end{align}
From this we conclude that a perfect rotation  around the axis $\tilde{x} = \cos(\phi) \sigma^x + \sin(\phi) \sigma^y$ can be performed if $\delta=0$ and $t = t_d=\frac{\pi}{2\A}$, meaning that each computational basis state is inverted into the other. Specifically, at $\phi=0$, we retrieve the $\sigma^x$ gate, and at $\phi = \pi/2$ the gate $\sigma^y$. We will use the term `resonant driving' to indicate the choice $\delta=0$ with the aim to implement such transitions between computational basis states (up to phases), and refer to $t_d$ as the driving time. 

We also consider what happens whenever the driving is off-resonant, in the limit where $|\delta| \gg |\A|$. Here, $\vec{n}$ points mainly in the $\sigma^z$ direction, causing hardly any mixing of the computational basis states, but rather giving the states a hard-to-predict relative phase. This effect is otherwise known as the Autler-Townes or AC Stark effect. 

\subsection{Driving in a many-body system}
For our purposes, we want to drive a many-body system with a single driving field of fixed frequency $\omega$.  However, each pair of eigenstates $s_1, s_2$ of $H_\text{bg}$ will have a different values of energy gap $\delta$ and a different $\A= \A' \bra{s_1} H_\text{drive} \ket{s_2}$, leading to different behaviour among each pair. However, we argue that it is possible to 
\begin{enumerate}
\item Incite a transition between two states that have a unique energy gap $\Delta E$, by tuning $\delta = 0$ for that pair of states.
\item Leave all other states approximately unchanged.
\end{enumerate}
Leaving all other states unchanged is particularly challenging, in part because $H_\text{bg}$ is responsible for a continuously growing dynamical phase on each eigenstate. There are various ways to regain control over these phases:
\begin{itemize}
\item One keeps track of all dynamical phases that occur throughout the whole protocol, and undoes these at the end of the resonant driving. With $2^N$ eigenstates, this is generically infeasible unless there is an exploitable symmetry between the eigenstates. An example of such a symmetry occurs in free particle Hamiltonians, where many-particle states have energies which are sums of single-particle energies. In such cases, an appropriate eigengate can map all accumulated phases back onto a local qubit, such that phases can be undone with a local phase shift on each of the qubits. 
\item One chooses the driving time $t_d$ (and hence the corresponding amplitude $\A$) precisely such that all two-level systems make an integer number of rotations during the protocol. This amounts to making sure that $\delta t_d = k 2\pi \ (k \in \mathbb{Z})$ for each transition. The choice of $t_d$ becomes increasingly constrained as the systems grows and increasingly many values of $\delta$ have to be taken into account. 
\item One decomposes the driving procedure in two steps, which cancel each other's accumulated phase to leading order. 
\end{itemize}
Throughout the rest of the paper, we proceed by using the latter approach, which we dub the halfway-inversion. We address it in the next subsection. 

As an aside, note the difference between driving of the form $\cos(\omega t) H_\text{drive}$ in Eq. $\ref{eqn:generaldriving}$ compared to $\exp( i \omega t ) \sigma^+ + \text{h.c.}$ in Eq. \ref{eqn:tlsdriving}. A subtlety is that the latter gives rise to \emph{exact} transitions upon driving on resonance, whereas the first gives an accurate inversion only after assuming the rotating wave approximation  \cite{Irish2005, Wu2007}. Throughout this paper, we assume this approximation ($\A \ll \omega$) to hold.

\subsection{The halfway-inversion}
Let us first remind the reader of the Hahn (or spin) echo. We consider a set of two-level systems which each evolve under a Hamiltonian of the form $H = \varepsilon_j Z$, where the energy $\varepsilon_j$ is unknown for each system $j$. For any time $t$, one can enforce each of these system to return to their initial state by implementing a Hahn echo, where the system is suddenly flipped by the $\sigma_x$ operation at times $t/2$ and $t$. This is easily seen to work for any choice of $\varepsilon_j$, by using $\sigma^x \sigma^z \sigma^x = -\sigma^z$:
\begin{align}
U_\text{tot} = ( \sigma^x e^{-i \varepsilon_j \sigma^z \frac{t}{2} } \sigma^x ) e^{-i \varepsilon_j \sigma^z \frac{t}{2} } = e^{+i \varepsilon_j \sigma^z \frac{t}{2} } e^{-i \varepsilon_j \sigma^z \frac{t}{2} } = \mathds{1}.
\label{eqn:hahnecho}
\end{align}
Note that during the driving step of our protocol, the term $H_\text{bg}$ of Eq. $\ref{eqn:generaldriving}$ causes its eigenstates to rotate according to their energies $\lambda_j$. If $H_\text{bg}$ can be obtained from some other Hermitean operator using an eigengate, then sequential application of two eigengates $U_\text{eg}^2$ precisely inverts the spectrum of $H_\text{bg}$, effectively applying $\sigma^x$ to each two-level system formed by a pair of eigenstates. Importantly, note the difference between the driving operator of the form of Eq. \ref{eqn:tlsdriving}, which swaps  \emph{only} the states that are on resonance but is error-prone and slow ($|\A| << |\delta|$), contrasted with $U_\text{eg}^2$ which inverts \emph{all} states, and is exact and relatively fast. 

Inspired by the Hahn echo, we propose to follow the same approach on a many-body system: 
\begin{align}
U_\text{tot} = \sigma^x \ U^{\text{(lab)}}_\text{drive}(\phi_2, t_d/2) \ \sigma^x \ U^{\text{(lab)}}_\text{drive}(\phi_1, t_d/2).
\label{eqn:hwp-steps}
\end{align}
Here, the form of $U^{\text{(lab)}}_\text{drive}$ was derived in Eq. \ref{eqn:Udrive}, but we stress that the earlier derivation was done in the \emph{rotating frame}. The operations $U_\text{tot}$ and $\sigma^x$, however, are stated in the lab frame. We can translate the whole equation back to the lab frame by using 
\begin{align*}
U_\text{drive}^\text{(\text{lab})}(\phi, t) = \exp \left( -i \omega \sigma^z t / 2 \right)\ U_\text{drive}^\text{(\text{rf})}(\phi,t),
\end{align*}
such that we obtain
\begin{align*}
U_\text{tot} &= \underbrace{ \sigma^x \  \exp \left( -i \omega \sigma^z \frac{t_d}{4} \right) \ U_\text{drive}^\text{(\text{rf})}(\phi_2,t_d/2)  \ \sigma^x \  \exp \left( -i \omega \sigma^z \frac{t_d}{4} \right)  }_{ I } 
\ U_\text{drive}^\text{(\text{rf})}(\phi_1,t_d/2).
\end{align*}
The first part can be rewritten as 
\begin{align*}
I =& \sigma^x  \ \exp \left( -i \omega t_d \sigma^z / 4 \right)  \ \exp \left[ -i t \ \vec{\sigma} \cdot \begin{pmatrix} n_x \\ n_y \\ n_z \end{pmatrix} \right] \sigma^x \exp \left(-i \omega t_d \sigma^z / 4 \right)  \\
= \ &  \sigma^x  \ \exp \left(-i \omega t_d \sigma^z / 4 \right)  \ \exp \left[ -i t \ \vec{\sigma} \cdot \begin{pmatrix} n_x \\ n_y \\ n_z \end{pmatrix} \right] \exp \left( +i \omega t_d \sigma^z / 4 \right) \sigma^x  \\
= \ & \sigma^x   \ \exp \left[ -i t \ \vec{\sigma} \cdot \begin{pmatrix} n_x \cos( \omega t_d / 2 ) + n_y \sin( \omega t_d / 2  ) \\ n_y \cos(\omega t_d / 2) - n_x \sin( \omega t_d / 2) \\ n_z \end{pmatrix} \right] \sigma^x  \\
= \ & \exp \left[ -i t \ \vec{\sigma} \cdot \begin{pmatrix}
\A \cos( \phi_2 + \omega t_d / 2 )  \\ - \A \sin( \phi_2 + \omega t_d / 2 ) \\ - \delta / 2 
\end{pmatrix} \right].
\end{align*}
In the last step, we used that $\vec{n} = ( \A \cos(\phi_2) , \A \sin(\phi_2) , \delta / 2 )^T$ and applied several trigonometric product-to-sum identities. All in all, we find that 
\begin{align*}
U_\text{tot} &= \exp \left[ -i \frac{t_d}{2} \ \vec{\sigma} \cdot \begin{pmatrix}
\A \cos( \phi_2 + \omega t_d/2 )  \\ - \A \sin( \phi_2 + \omega t_d / 2 ) \\ - \delta / 2 
\end{pmatrix} \right]  \exp \left[ -i \frac{t_d}{2} \ \vec{\sigma} \cdot \begin{pmatrix}
\A \cos( \phi_1 )  \\  \A \sin( \phi_1 ) \\  \delta / 2 
\end{pmatrix} \right].
\end{align*}
Let us first show that, despite the halfway inversion, it is still possible to perform a transition without any error if $\delta = 0$. Note that, in this picture, we perform two consecutive $\pi/2$ pulses, which form an optimal $\pi$-pulse if and only if both rotation axes align. To this end, we fix
\begin{align}
\phi_2 = - \phi_1 - \omega t_d /2 ,
\label{eqn:phi2}
\end{align}
such that the overall rotation becomes 
\begin{align}
U_\text{tot} &= \exp \left[ -i \frac{t_d}{2} \ \vec{\sigma} \cdot \begin{pmatrix}
\A \cos( \phi_1 )  \\  \A \sin( \phi_1 ) \\ - \delta / 2 
\end{pmatrix} \right]  \exp \left[ -i \frac{t_d}{2} \ \vec{\sigma} \cdot \begin{pmatrix}
\A \cos( \phi_1 )  \\  \A \sin( \phi_1 ) \\  \delta / 2 
\end{pmatrix} \right].
\label{eqn:utothwp}
\end{align}
It is now clear that, for $\delta = 0$, any rotation axis in the $x-y$ plane can be obtained by an appropriate choice of $\phi_1$. We sketch a more intuitive picture of what happens to an inverting state in this case in Appendix \ref{app:hwp-intuition}.
On the other hand, a rotation around \emph{only} $\sigma^z$ is impossible as the two driving steps would cancel. We will shortly treat off-resonant rotations which are only approximately around $\sigma^z$. For now, note that we could have chosen a different flip $\sigma^x$ together with rotation vector $\vec{n}$ precisely such that \emph{all} rotations would cancel, resulting again in the operation $U_\text{tot} = \mathds{1}$ as in Eq. \ref{eqn:hahnecho}. However, this prohibits us from performing the required inversion on the resonant pair of states. The impossibility to cancel the rotations induced by the $\sigma^x$ and $\sigma^y$ components is our main source of errors. 

Let us now consider the accuracy of $U_\text{tot}$ in the case of off-resonance, $\A \ll \delta$, where we aim to not cause any transitions at all. We define the error $\mathcal{E}$ of a unitary $U$ with respect to a target unitary $U_\text{goal}$ as
\begin{align}
\mathcal{E}(U, U_\text{goal}) = 1 - \frac{1}{\text{dim}(U)}  | \text{Tr}( U U_\text{goal}^\dagger ) |.
\end{align}
Comparing $U_\text{tot}$ to the identity operator for the case of Eq. \ref{eqn:utothwp}, we find 
\begin{align}
\mathcal{E}(U_\text{tot}, \mathds{1}) &= 1 - | \cos^2(n t_d /2) - \frac{(\A^2 - \frac{\delta^2}{4})}{n^2} \sin^2(n t_d/2) | \nonumber \\
&= 1 -| 1 - \sin^2(n t_d / 2) \left( \frac{\A^2}{\A^2 + \frac{\delta^2}{4}} \right) | \nonumber \\
&= \sin(n t_d /2 )^2 \left(  \frac{ 8 \A^2}{\delta^2} + O\left( \frac{\A^4}{\delta^4} \right) \right).
\label{eqn:hwp_error}
\end{align}
This shows that the error can be made arbitrarily small, by choosing a smaller $\A/\delta$, or equivalently, a longer gate time $t$ while keeping $H_\text{bg}$ constant.

The factor $\sin(nt_d/2)$ could in principle cause the error to vanish if $n t_d = k \pi ~ (k \in \mathbb{Z})$. Note that with many two-level systems, this is highly unlikely to happen and hard to track. Interestingly, by just considering the energy gaps $\delta$, one can shave off another two orders of $\frac{\A}{\delta}$ from $\mathcal{E}$ in the specific case where $\delta t_d = k 4 \pi ~ (k \in \mathbb{Z})$:
\begin{align*}
n t_d / 2 =&   \sqrt{\A^2 + \frac{\delta^2 }{4} } t_d / 2 \\ 
= & \frac{\delta t_d}{4}  \left[ 1 + 2 \cdot  \frac{4 A^2}{\delta^2} + O \left( \frac{A^4}{\delta^4} \right) \right]  \\
= & k \pi + \frac{2 \pi \A}{\delta} +  O \left( \frac{\A^2}{\delta^2} \right)  \\
\sin( n t_d / 2 )^2 = & \frac{4 \pi^2 \A^2}{\delta^2} +  O\left( \frac{\A^4}{\delta^4} \right) && \text{if } ~ k \in \mathbb{Z}.
\end{align*}
Hence, in the special case that all dynamic phases due to $H_\text{bg}$ reset, $\mathcal{E}(U_\text{tot}, \mathds{1}) =  O\left( \frac{\A^4}{\delta^4} \right)$. 

Unfortunately, we do not find the same $O\left( \frac{\A^4}{\delta^4} \right)$ scaling when driving a many-body system, even when engineering the energy gaps $\delta$. Numerically, we find the culprit to be the two-level systems consisting of one spectator state and one transitioning state: the off-resonant transition between these pairs is not accounted for by the halfway inversion, and hence still contributes an error of the order $O\left( \frac{\A^2}{\delta^2} \right)$. Nonetheless, the cases where $\delta t_d \approx k 4 \pi$ lead to a significant improvement of our protocol's fidelity, as we will see in the numerical results in Sec. \ref{sec:numerics}.

\subsection{Putting it all together: Resonantly driven multi-qubit gates}
\label{sec:resdriveprotocol}
We turn back to the many-body Hamiltonian proposed at the start of this chapter,
\begin{align*}
H = H_\text{bg} + \A' \cos(\omega t + \phi) H_\text{\text{drive}}
\end{align*}
and its resulting unitary evolution $U_\text{drive}(\phi, t)$. We found that for sufficiently small $\A'$ and an appropriately chosen frequency $\omega$ and driving time $t_d$, we asymptotically approximate the operation $\iswap$ which selectively exchanges two basis states:
\begin{align*}
\iswap_{t_1,t_2} :=& -i e^{i \phi} \ket{t_1} \bra{t_2}  -i e^{-i \phi } \ket{t_2}\bra{t_1} +  \sum_{j \not\in \{ t_1, t_2 \} }   \ket{j}\bra{j}.
\end{align*}
Note the difference between the phases $e^{\pm i \phi}$: we used the convention that $\ket{t_2}$ is the state with the \emph{lower} energy (e.g. $Z \ket{t_2} = - \ket{t_2}$ in the two-level system formed by $\ket{t_1}$, $\ket{t_2}$). This gate is implemented by the sequence 
\begin{align*}
\iswap_{t_1,t_2} &\approx U_\text{eg}^\dagger ~ U_{\text{drive}}\left(-\phi - \frac{\omega t_d}{2}, \frac{t_d}{2}\right) ~ U_\text{eg}^2 ~  U_{\text{drive}}\left(\phi, \frac{t_d}{2}\right) ~ U_\text{eg}  && \text{(with halfway inversion)} 
\end{align*}

Alternatively, one can leave out the halfway-inversion but undo the dynamical phases by inverting the spectrum of $H_\text{bg}$, optionally even in the computational basis after an eigengate is performed:
\begin{align*}
\iswap_{t_1,t_2} & \approx U_\text{eg} e^{+i H_\text{bg} t_d}  U_{\text{drive}}(\phi, t_d)  U^\dagger_\text{eg}   && \text{(without halfway inversion)}\\
& \approx e^{+i H_\text{cb} t_d} U_\text{eg} U_{\text{drive}}(\phi, t_d)  U^\dagger_\text{eg}  
\end{align*}
where $H_\text{cb} = U_\text{eg} H_\text{bg} U_\text{eg}^\dagger$ is the eigengate-partner of $H_\text{bg}$ in the computational basis. The phases of the $\iswap$ operation are again $-i \exp(\pm i \phi)$, the same as with the halfway inversion.

\section{Polychronakos' model}
\label{sec:pc}

We now turn to a concrete model which provides us a $H_\text{bg}$ in which resonant driving can be performed, and for which a map between the eigenbasis and computational basis can be found. The model we consider was first described by Polychronakos \cite{Polychronakos1993}, but we will follow the definitions of Frahm \cite{Frahm1993}, who found an associated algebraic structure which we employ in our protocol. This algebraic structure is similar in spirit to the celebrated Yangian symmetry of Haldane-Shastry model \cite{Haldane1988,Shastry1988,Haldane1992}, to which the Polychronakos chain is a close relative. Both models are members of a wider class of integrable 1D systems with inverse-square two-body interactions, going back to the Calogero-Moser-Sutherland model of interacting particles on a line \cite{Calogero1969, Calogero1969a,  Calogero1971, Sutherland1971, Sutherland1972, Sutherland1975, Moser1975}.

We consider a one-dimensional chain of $N$ spin-$\frac{1}{2}$ particles, with particle $j$ fixed at position $x_j$, evolving under the Hamiltonian 
\begin{align*}
H_\text{P} &= \sum_{j<k} h_{jk} P_{jk}, \quad  \text{where \ } h_{jk} = \frac{1}{(x_j - x_k)^2}, \\
P_{jk} &= \frac{1}{2} \left( \mathds{1}_{j} \mathds{1}_{k} + \sum_{\alpha = {x,y,z}} \sigma^\alpha_j \sigma^\alpha_k \right) 
= \frac{1}{2} \begin{pmatrix}
1 & 0 & 0 & 0 \\
0 & 0 & 1 & 0 \\
0 & 1 & 0 & 0 \\
0 & 0 & 0 & 1 \\
\end{pmatrix}.
\end{align*}  
The locations $x_j$ are given by the equilibrium positions of the classical Colagero system with potential 
\begin{align*}
V(x_1, ..., x_N) = \frac{1}{2} \sum_j x^2_j + \sum_{j<k} \frac{1}{(x_j - x_k)^2}
\end{align*}
or, equivalently, by the roots of the Hermite polynomial $H_N(x)$. Frahm was able to describe the eigenbasis by finding ladder operators, and in particular, defined the following operators:
\begin{align*}
L_0^\alpha &= \frac{1}{2} \sum_{j=1}^{N} x_j \vec{\sigma}_j^\alpha  		&& \alpha, \beta, \gamma  \in \{ x,y,z \}\\
L_1^\alpha &= \frac{1}{4} \sum_{j \neq k} w_{jk} \epsilon^{\alpha \beta \gamma} \vec{\sigma}_j^\beta \vec{\sigma}_k^\gamma 	&& w_{jk} = \frac{1}{x_j - x_k}  
\end{align*}
where $\epsilon^{\alpha \beta \gamma}$ is the Levi-Civita symbol or fully anti-symmetric tensor. These operators were found to have the following relation with $H_\text{P}$:
  \begin{align*}
  [ H_\text{P},  L_0^\alpha ] &= i L_1^\alpha \\
  [ H_\text{P},  L_1^\alpha ] &= -i L_0^\alpha.
  \end{align*}

\subsection{Mapping between eigenbases}
\begin{figure}[t]
\centering
\includegraphics[width=.55\textwidth]{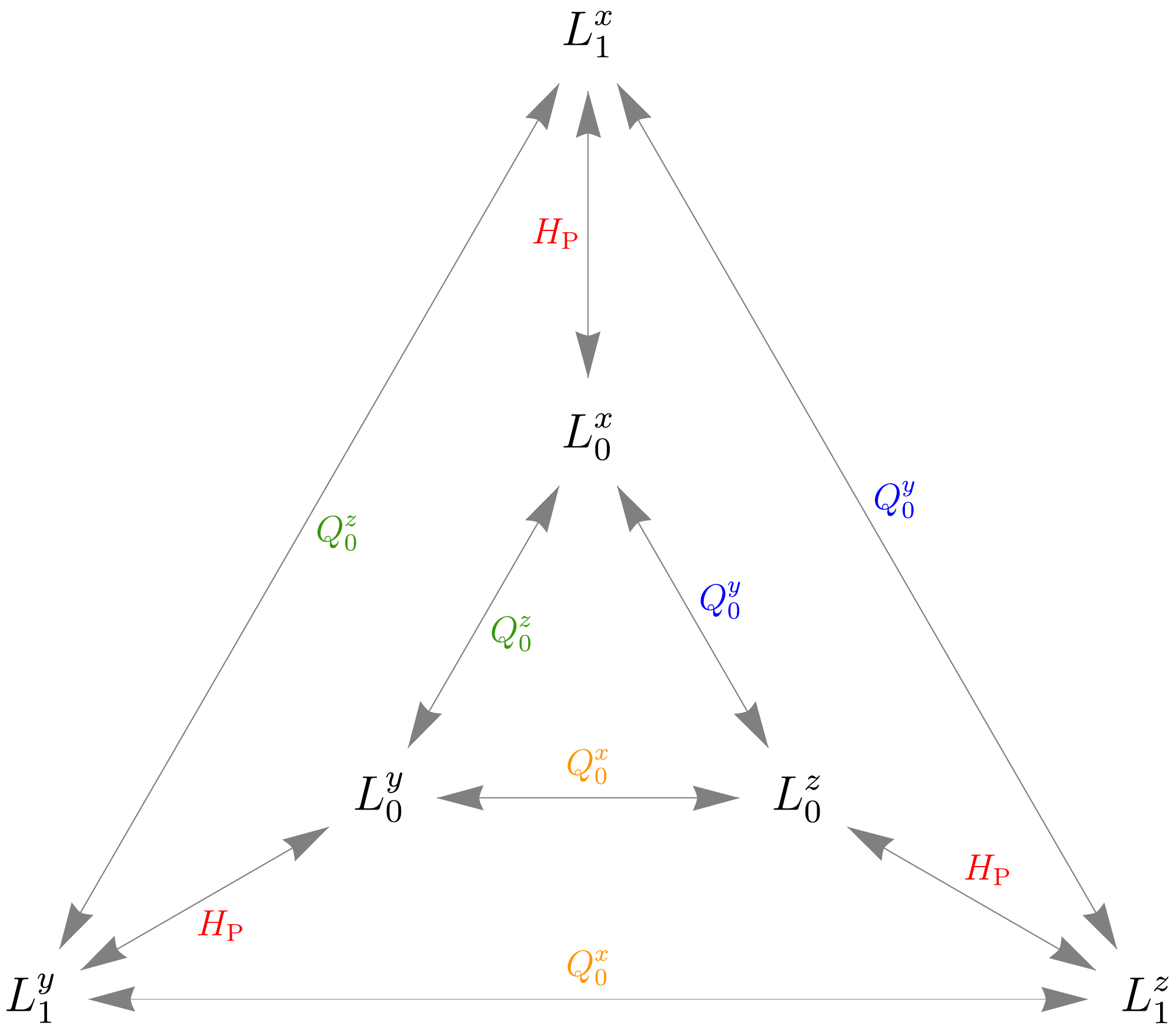}
\caption{A map of various eigengates between the operators $\{ L_r^\alpha \}_{r,\alpha}$ in Polychronakos' model. A quench by the operator next to an arrow implements the corresponding eigengate. }
\label{fig:polychronakos_egs}
\end{figure}
By noting that the eigenstates of $L_0^z$ are the computational basis states, and the commutation relations between $( L_0^\alpha, L_1^\alpha, H_P )$ are of the form of Eq. \ref{eqn:alt_comm}, we readily obtain two methods to obtain an eigengate for $L_1^z$: either by continuous evolution 
\begin{align*}
U_\text{eg}^P = \exp( -i H_\text{P} \pi / 2 ),
\end{align*}
or by adiabatically evolving $H_\text{adiabatic} = \cos( t ) L_0^z + \sin( t ) L_1^z$ for $t\in [0,\pi/2]$. Interestingly, owing to the form of $L_0^z$, applying the operation $U_\text{eg}^P$ twice performs a spatial mirror inversion and perfect state transfer on the spin chain \cite{Bose2007}. 

Another eigengate which interchanges the operator superscripts $x$, $y$ and $z$ can be formed by quenching with the total spin operator
\begin{align*}
Q_0^\beta &= \frac{1}{2} \sum_j \sigma_j^\beta  && \beta \in \{ x,y,z \} 
\end{align*}
as it satisfies 
\begin{align*}
[ Q_0^\alpha , L_r^\beta ] &= i \epsilon^{\alpha \beta \gamma} L_r^\gamma. && r \in \{ 0, 1 \}.
\end{align*}
The different eigengates in this model are summarized in Fig. \ref{fig:polychronakos_egs}.

\subsection{Resonant driving in Polychronakos' model}
\begin{figure}[t]
\centering
\includegraphics[width=.7\textwidth]{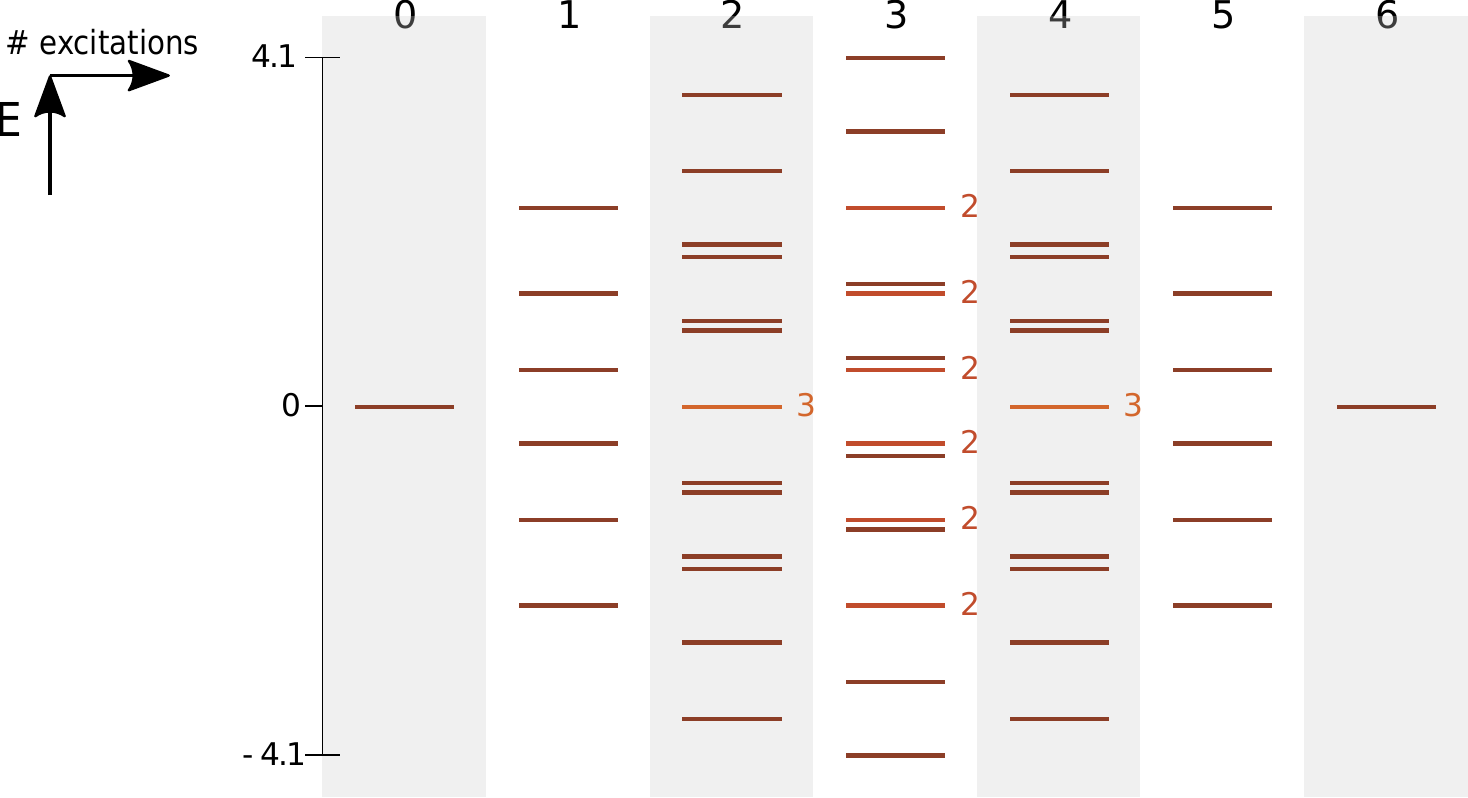}
\caption{The spectrum of each of the operators $\{ L_i^\alpha \}$, depicted for $N=6$. The horizontal ordering denotes the number of $\alpha$-excitations (or Hamming weight) $\frac{N}{2} - < Q_0^\alpha>$.  Subspaces with multiplicity larger than 1 have their multiplicity displayed to their right.}
\label{fig:L0spec}
\end{figure}
The energy spectra of all $\{ L_r^\alpha \} $ are identical to that of $L_0^z$ as these operators are linked by an isospectral transformation. Since $H_P$ commutes with each total spin operator $\{ Q_0^\alpha \}$, we conclude that the eigensystem of each $L_i^\alpha$ separates into non-interacting blocks of constant $Q_0^\alpha$ (e.g. total spin projection in the $\alpha$ direction). From here onwards, we will focus on $\alpha = z$, but we stress that identical results hold for the $x$ and $y$ superscripts, up to a local basis transformation. 

The spectrum of $L_0^z$ for $N=6$ is depicted in Fig. \ref{fig:L0spec} with energies represented vertically and the value of $Q_0^z$ horizontally. Let $\ket{ \{ k_1, k_2, \ldots, k_p \} }$ with $k_1 < k_2 < \ldots < k_p$ represent the state with qubits $k_j$ in state $\ket{1}$ and all other qubits in the state $\ket{0}$. The energies of states expressed in this notation are conveniently calculated as 
\begin{align*}
L_0^z \ket{ \{ k_1, k_2, \ldots, k_p \} } = \sum_{j=1}^p x_{k_j} \ket{ \{ k_1, k_2, \ldots, k_p \} }.
\end{align*}
In words: for each qubit in state $\ket{1}$, add energy equal to the position $x_j$ of that qubit. 

Because the positions $x_j$ are symmetric around $0$, the highest- and lowest energy states are nondegenerate for even $N$, and are given by:
\begin{align*}
\ket{\text{high}} &= \ket{t_1} = \ket{ 0 }^{ \otimes \frac{N}{2} } \ket{ 1 }^{ \otimes \frac{N}{2} } = \ket{ \{ \frac{N}{2} + 1, \ldots, N \} } && (N \text{ even}) \\
\ket{\text{low}} &= \ket{t_2} = \ket{ 1 }^{ \otimes \frac{N}{2} } \ket{ 0 }^{ \otimes \frac{N}{2} } = \ket{ \{ 1, \ldots, \frac{N}{2} \} }
\end{align*}
As the energy gap between $\ket{t_1}$ and $\ket{t_2}$ is unique, we may select these two states for a resonantly driven multi-qubit gate. However, these states are of product form in the eigenbasis, hence a local operator cannot have nonzero matrix element between $\ket{t_1}$ and $\ket{t_2}$. Therefore, we employ an eigengate $U^P_{\text{eg}}$ to turn $\ket{t_1}, \ket{t_2}$ into spatially extended states $\ket{t_1}_{L_1^z}$ and $\ket{t_2}_{L_1^z}$, while preserving the spectrum and in particular the unique energy gap. We can then resort to the driving protocol proposed in Sec. \ref{sec:resdriveprotocol} to create an $\iswap_{t_1,t_2}$ gate by choosing $H_\text{bg} = L_1^{z}$ and choosing for $H_\text{drive}$ any operator that couples $\ket{t_1}_{L_1^z}$ and $\ket{t_2}_{L_1^z}$. 

It is not clear in general what choices of $H_\text{drive}$ lead to lower gate errors at similar driving times. One constraint is that the coupling must preserve the expectation value of $Q_0^\alpha$ (i.e. the number of spins in state $\ket{1}$), indicating that couplings such as $\sigma^x$ or $\sigma^z \otimes \sigma^y$ cannot drive the required transition. For some common nontrivial 1- and 2-local driving operators and small system sizes $N$, we tabulate the matrix elements $_{L_1^z} \bra{t_1} H_\text{drive} \ket{t_2}_{L_1^z}$ below. 

\begin{center}
  \begin{tabular}{ | l | c | }
  	\hline
    \multicolumn{2}{ | c | }{$N=4$}  \\ \hline
	$H_\text{drive}$ 			& $ _{L_1^z} \bra{t_1} H_\text{drive} \ket{t_2}_{L_1^z} $  \\ \hline
	$\sigma^z_2$				& $-0.413049 i$ \\
    $\sigma^z_2 \sigma^z_3$ 	& $0.829345$ \\
    $\sigma^z_1 \sigma^z_4$ 	& $0.829345$ \\
   	$\sigma^x_2 \sigma^x_3$ 	& $-0.552743$ \\
	$\sigma^y_2 \sigma^y_3$ 	& $-0.552743$ \\
    $\sigma^x_1 \sigma^x_4$ 	& $0.390066$ \\
	$\sigma^x_2 \sigma^y_3$ 	& $0$ \\
	\hline
  \end{tabular}
\quad
  \begin{tabular}{ | l | c | }
  	\hline
    \multicolumn{2}{ | c | }{$N=6$}  \\ \hline
	$H_\text{drive}$ 			& $ _{L_1^z} \bra{t_1} H_\text{drive} \ket{t_2}_{L_1^z} $  \\ \hline
	$\sigma^z_3$				& $0.116012 i$ \\
    $\sigma^z_3 \sigma^z_4$ 	& $-0.327919$ \\
    $\sigma^z_1 \sigma^z_5$ 	& $-0.353636$ \\
    $\sigma^z_1 \sigma^z_5$ 	& $0.265128$ \\ 
	$\sigma^x_3 \sigma^x_4$ 	& $0.200378$ \\
	$\sigma^x_2 \sigma^x_3$ 	& $-0.147838$ \\
	$\sigma^x_1 \sigma^x_5$ 	& $-0.14341$ \\
	\hline
  \end{tabular}
\quad
  \begin{tabular}{ | l | c | }
  	\hline
    \multicolumn{2}{ | c | }{$N=8$}  \\ \hline
	$H_\text{drive}$ 			& $ _{L_1^z} \bra{t_1} H_\text{drive} \ket{t_2}_{L_1^z} $  \\ \hline
	$\sigma^z_4$				& $ -0.027894 i $ \\
    $\sigma^z_4 \sigma^z_5$ 	& $ -0.0839009 $ \\
	$\sigma^z_1 \sigma^z_6$ 	& $ 0.120287 $ \\
	$\sigma^z_2 \sigma^z_7$ 	& $ 0.131574 $ \\
	$\sigma^x_4 \sigma^x_5$ 	& $ 0.0471167 $ \\
	$\sigma^x_1 \sigma^x_6$ 	& $ 0.0502561 $ \\
	$\sigma^x_2 \sigma^x_7$ 	& $ 0.0452589 $ \\
	\hline
  \end{tabular}
\end{center}

\subsection{Tracking dynamical phases}
As the energies of $L_1^z$ are sums of single-excitation energies, it is possible to keep track of dynamical phases of individual states efficiently. One could in principle perform an eigengate $U^P_{eg}$, drive a transition in time $t_d$, and map back to the computational basis using $(U^P_\text{eg})^\dagger$. The accumulated dynamical phases on qubit $j$ is then equal to $x_j t_d$, which may be undone by a single-qubit phase gate, or by following the halfway inversion protocol.

\section{Numerical results}
\label{sec:numerics}

We test our claims by simulating the driving step of our protocol, through numerically solving Schr\"{o}dinger's equation given by the Hamiltonian
\begin{align*}
H_P(t) =  L_1^z + \A_P \cos(\omega t) H_\text{drive}.
\end{align*}
We consider the cases $N=4$ and $N=6$, and two different driving operators $H_\text{drive}$ which fit in the connectivity of the linear chain. The driving frequency $\omega$ is always chosen to be exactly the energy gap between states $\ket{t_1}_{L_1^z}$ and $\ket{t_2}_{L_1^z}$. Moreover, after fixing the driving time $t_d$, we choose $\A_P$ such that a $\pi$-rotation occurs between the transitioning states, e.g. $|  _{L_1^z}\bra{t_2} H_\text{drive} \ket{t_1}_{L_1^z} | ~ \A_P ~ t_d = \pi$. Apart from the halfway-inversion, we apply no further optimizations to the protocol. 

\begin{figure}[p]
\centering
\includegraphics[width=.7\textwidth]{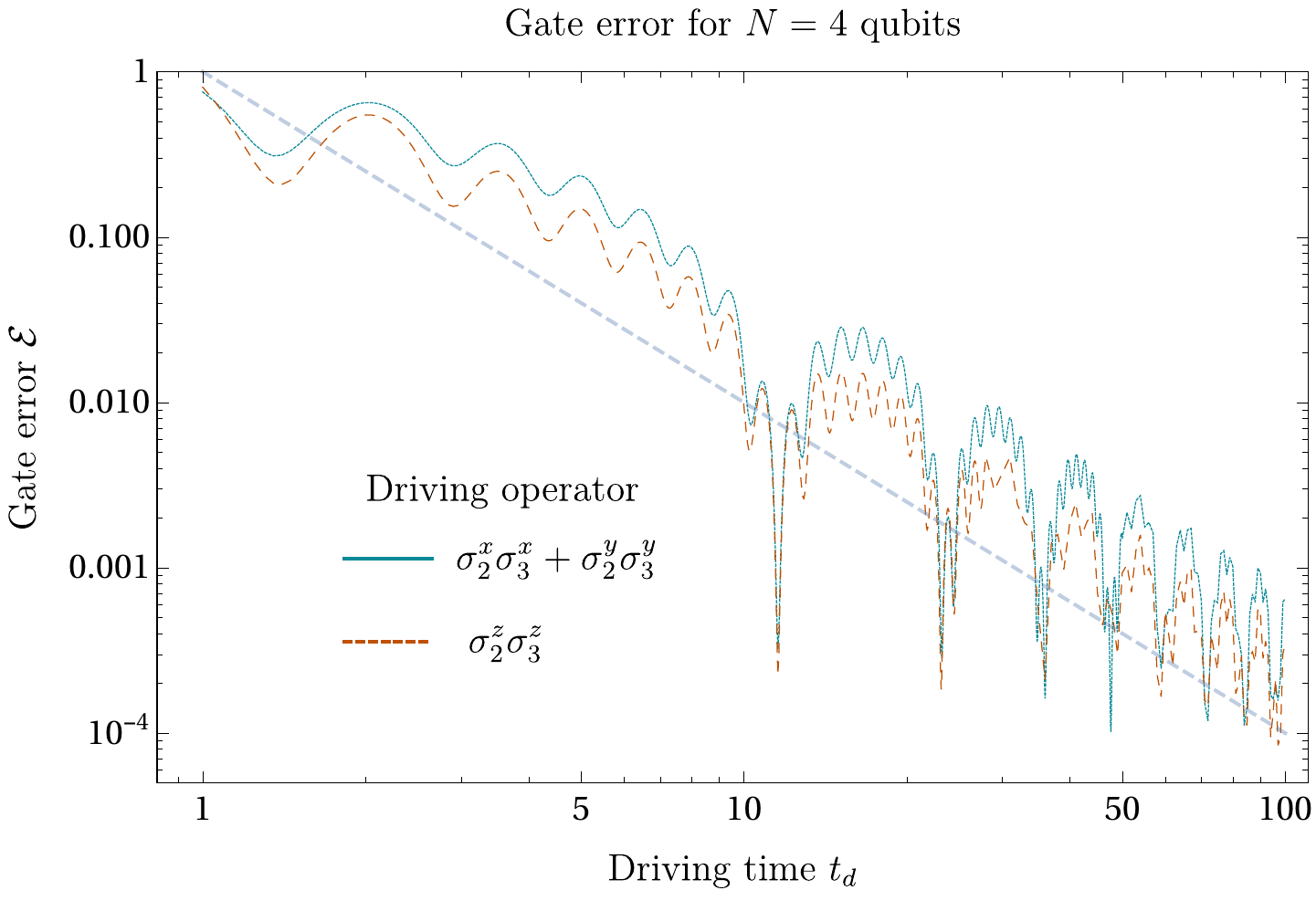} 
\includegraphics[width=.7\textwidth]{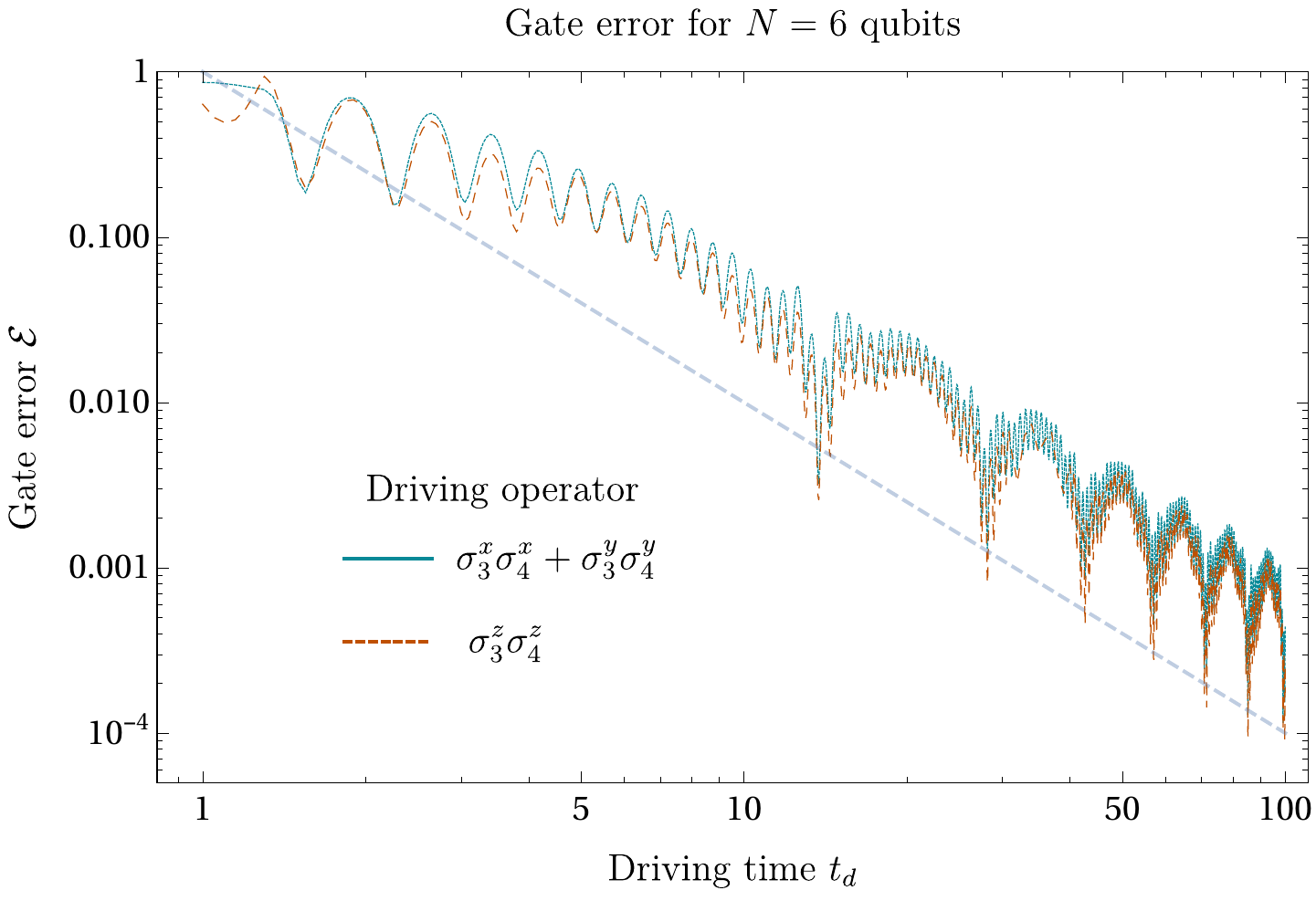}
\caption{Gate error due to the resonant driving stage of our protocol, given various gate times and two different choices of $H_\text{drive}$, for a number of qubits equal to $N=4$ (top) and $N=6$ (bottom). In these results, we applied the halfway inversion, but no further optimizations. The dashed line follows $\mathcal{E} = t_d^{-2}$. Although the fidelity is strongly oscillatory in $t_d$, a global tendency towards inverse quadratic decay is clearly visible.} 
\label{fig:drivefids}
\end{figure}

The results are presented in Fig. \ref{fig:drivefids}. For extremely short driving times, where $t_d \approx 1$ such that $\A_P$ is of the order of energy differences of the background Hamiltonian, the gate is highly inaccurate. However, for longer driving times, $t_d > 10$, the gate becomes increasingly accurate, with the error decaying roughly as $t_d^{-2}$ as expected. We also note that the fidelity seems to not strongly depend on the choice of driving operator. 

\begin{figure}
\centering
\includegraphics[width=.7\textwidth]{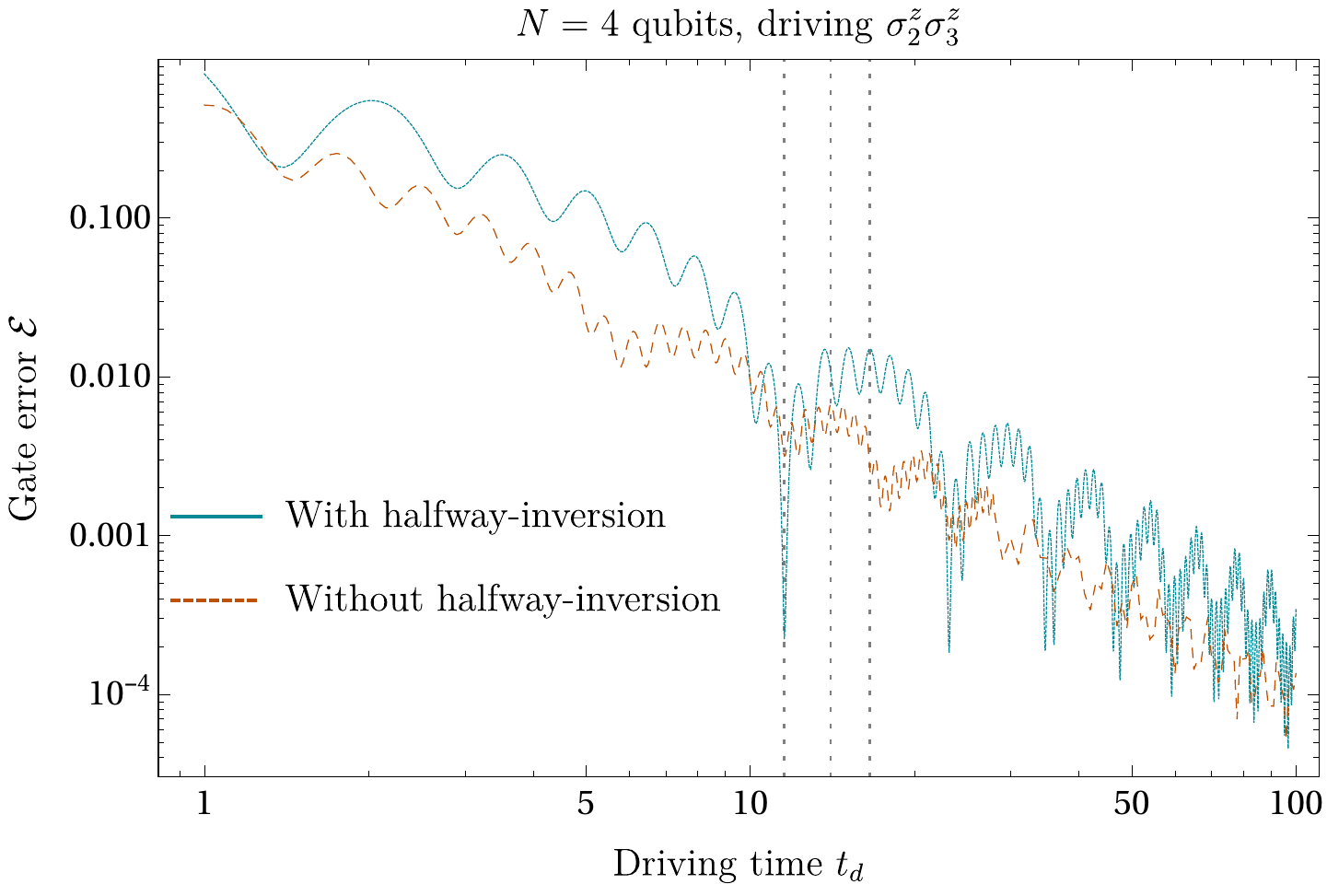} 
\caption{A comparison of the gate error either with or without a halfway inversion applied, for the case $N=4$ and $H_\text{drive} = \sigma_2^z \sigma_3^z$. The Halfway inversion shows stronger oscillatory behaviour, leading to minima that improve the total protocol fidelity by more than an order of magnitude at equal driving times. The times $t_d \in \{ 11.55, 14.05, 16.55 \}$ are highlighted with a gray, dotted line.}
\label{fig:compare_hwp}
\end{figure}

In Fig. \ref{fig:compare_hwp}, we depict the effect of the halfway-inversion compared to leaving it out. In the latter case, we undo the accumulated dynamical phases with the operation $\exp( +i L^z_0 t_d)$ after the system is mapped back to the computational basis. The graph shows that the halfway-inversion does not necessarily reduce the error at all times, but dramatically improves the error at very specific times. 

\begin{figure}
\centering
\includegraphics[width=.9\textwidth]{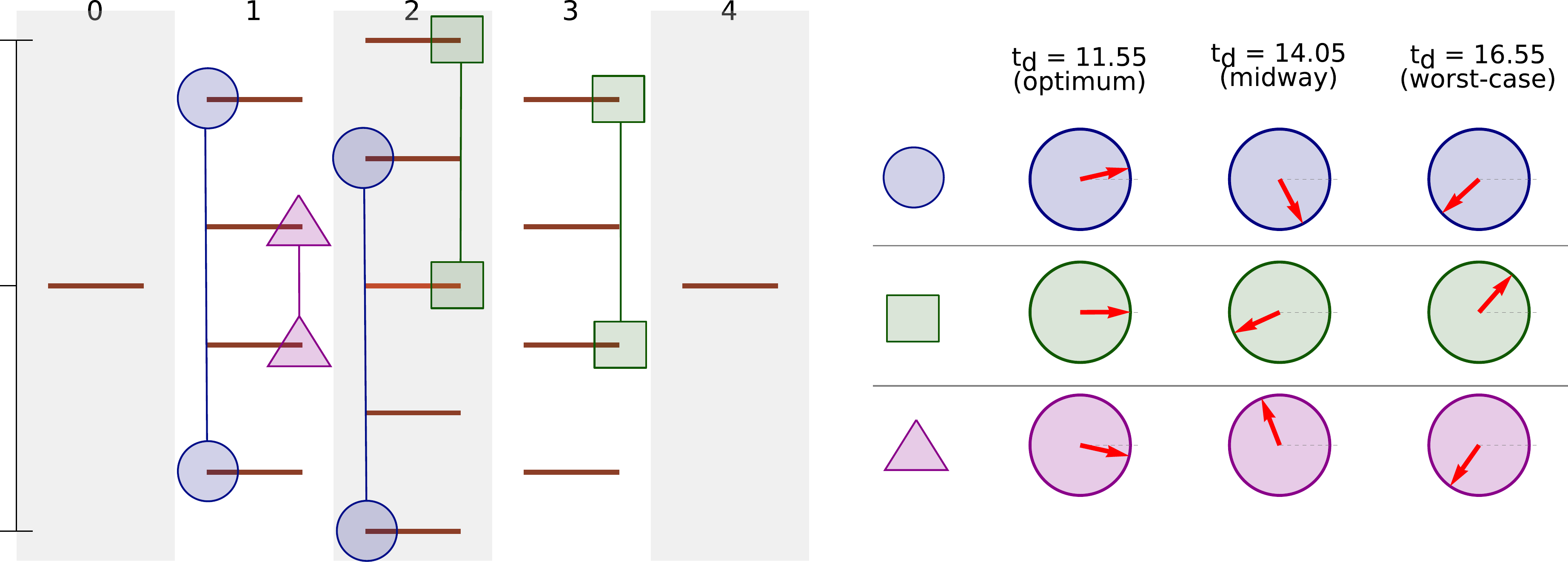}
\caption{For near-resonant two-level systems, the accumulated phases at the time of the halfway-pulse ($t_d / 2$) affect the fidelity of protocol. In the spectrum of $L_1^z$ with $N=4$ qubits, we select three different energy gaps (indicated by a pair of connected circles, squares or triangles), for which we depict the corresponding relative phases at $t_d$ as vectors on the unit sphere. A relative phase of $0$ corresponds to a vector pointing to the right (dashed lines). At $t_d = 11.55$, corresponding to a local minimum of our gate error $\mathcal{E}$, all phases are close to zero. On the other hand, at local maximum $t_d = 16.55$, the phases are all far from zero. The (non-)alignment of these phases explains the oscillatory behaviour of our protocol's error as a function of time. }
\label{fig:drivephases}
\end{figure}

We suspect that these specific times are precisely the times where, at the time of the halfway inversion, the relative phases of each two-level system are roughly $0$, causing the error's leading order term $8 \sin(nt) (\A^2 / \delta^2)$ (Eq. \ref{eqn:hwp_error}) to be minimized. We informally check this statement in Fig. \ref{fig:drivephases}, where the phases corresponding to highly optimal time $t_d = 11.55$ and local maximum $t_d = 16.55$, as well as the point precisely in between, are compared. The circles show the accumulated phase due to $H_\text{bg}$ for the indicated two-level system, with the rightmost point of the circle corresponding to zero phase. Clearly, the optimal timing is associated with near-optimal phase resets, whereas the more erroneous timing shows phases that could contribute a significant error of order $\A^2 / \delta^2$.

\subsection{Comparison of gate times}
So how do our driving times $t_d$ compare to other quantum gate times in the same system? A conventional two-qubit gate could be constructed if each of the individual Hamiltonian terms $h_{jk} P_{jk}$ could be turned on independently. A $\pi/4$ pulse would suffice to create maximal entanglement, so we require times $t$ such that $h_{jk} t = \pi / 4$. For $N=4$, the corresponding times $t$ lie between $0.86$ and $1.0$ for neighbouring qubits, and up to $8.56$ for the most distant qubits. These should be compared to the $11.55$ units of time required to perform an $\iswap_{1100, 0011}$ at low  error $\mathcal{E} < 0.001$. Hence, within the time of our four-qubit $\iswap$ operation, up to $13$ two-qubit gates could be done. 

Similarly, for $N=6$, neighbouring qubits could be entangled in times between $0.60$ and $0.81$, or up to $17.36$ to entangle the outermost qubits. This should be compared to the driving time $t_d = 13.75$ to obtain a driven gate with error $\mathcal{E} < 0.003$. Hence, our six-qubit resonant gate takes time equivalent to up to $23$ two-qubit gates. 

In general, it is unclear how to compare gate times between different gate sets, or how to optimally decompose $\iswap$ operations into smaller constituents. Turning to the well-studied Toffoli gate, the best bounds we could find are listed in Ref. \cite{Shende2009a}, stating that a Toffoli gate on four qubits requires between $8$ and $14$ CNOT operations. Note that these results assume full connectivity between all qubits, and don't account for the cost of single-qubit gates. For larger numbers of qubits, the CNOT cost is found to scale linearly in $N$ as long as auxiliary qubits may be used - without auxiliaries, it would be quadratic. Different physical interactions may also lead to different gate counts. For example, Ref. \cite{Schuch2003} finds that constructing a CNOT out of our interaction $P_{jk}$ requires two $\pi/4$ pulses, and constructing a mere three-qubit Toffoli using the closely related XY interaction requires as much as ten fundamental entangling operations. We conclude that it is not possible to make a rigorous comparison between different gate sets, but that the duration of our resonantly driven gate is competitive with conventional decompositions, with both approaches having unique advantages and disadvantages depending on the specific implementation.

\section{Discussion and summary of results}
\label{sec:discussion}
Let us turn back to the checklist presented in the introduction:

  \begin{checklist}
  \item[\done] \emph{A constantly applied background Hamiltonian $H_{\text{bg}}$ which has a unique energy gap between two eigenstates $\ket{t_1}_{H_\text{bg}}$ and $\ket{t_2}_{H_\text{bg}}$.} 
  \end{checklist}
  
In this work, we used that both the ground state and the highest energy state of $L^z_1$ are unique, hence constituting a unique transition. This condition is automatically fulfilled if the system is non-interacting and single-particle energies are non-degenerate. Other Hamiltonians that do not satisfy this sufficient requirement may need manual inspection of their spectrum. 

\begin{checklist}
  \item[\done] \emph{A driving field $H_{\text{drive}}$ which couples the states $\ket{t_1}_{H_\text{bg}}$ and $\ket{t_2}_{H_\text{bg}}$, whose amplitude may take the form of a cosine with appropriate frequency. }
\end{checklist}

We found that various 1- and 2-local operators were able to couple eigenstates of $L^z_1$. We did not address how these oscillatory driving fields could be physically implemented, which requires specialization to a specific experimental platform. 
  
\begin{checklist}
  \item[\done] \emph{An operation which maps (any) two computational basis states, call them $\ket{t_1}$ and $\ket{t_2}$, to energy eigenstates $\ket{t_1}_{H_\text{bg}}$ and $\ket{t_2}_{H_\text{bg}}$ respectively. We also need the inverse of this operation. }
\end{checklist}

We introduced the concept of an \emph{eigengate} which implements a basis transformation between two Hermitian operators. Although not every $H_\text{bg}$ may have a reasonable eigengate, we did find satisfying examples in Polychronakos' model, as well as in a different spin chain model in a previous work \cite{Groenland2018}. 

The eigengates considered here have the property that the \emph{whole} eigenbasis undergoes the correct map, but we stress that more general results may be obtained in which \emph{only} $\ket{t_1}$ and $\ket{t_2}$ map to the correct eigenstates. 

\begin{checklist}
  \item[\done]  \emph{An efficient method to keep track of the dynamical phases due to $H_{\text{bg}}$. } 
\end{checklist}
In an earlier work \cite{Groenland2018}, we proposed to tune the Hamiltonian such that all energies were integer multiples of each other, such that all dynamical phases reset after a known time. In this work, we note that for non-interacting many-body systems, the dynamical phases can be efficiently tracked, and may be undone by parallel single-qubit gates if all $H_{\text{bg}}$ eigenstates can be mapped back to the computational basis. Moreover, the halfway inversion, which maps each state with energy $E_j$ to a state with energy $-E_j$ halfway through the protocol, makes sure the dynamical phases cancel.  This operation can always be implemented by applying an eigengate for $H_\text{bg}$ twice. 

Taking all this together, the result is a highly nonlocal and nontrivial quantum operation formed by a continuous evolution of a Hamiltonian consisting of \emph{local} operators. A major weakness of this construction is the scaling with larger system sizes: For increasing $N$, the matrix elements $\ _{H_\text{bg}}\bra{t_2} H_\text{drive} \ket{t_1}_{H_\text{bg}}$ become increasingly small, causing the gate time to increase substantially. Therefore, the protocol seems to be competitive with conventional gate decompositions only for intermediate system sizes. It is an interesting open problem to find quantum systems for which these matrix elements remain constant or decay linearly at worst, while still satisfying the other requirements.

\section{Conclusion}
\label{sec:conclusion}

We presented a method to construct interesting gates on multiple qubits through resonant driving, where the resonance selects a unique two-dimensional subspace in which an inversion occurs. The error of the driving stage scales favourably with driving time as $\mathcal{E} \propto t_d^{-2}$, but may increase quickly as a function of the number of qubits $N$, due to the matrix elements $\ _{H_\text{bg}}\bra{t_2} H_\text{drive} \ket{t_1}_{H_\text{bg}}$ becoming increasingly small. Nonetheless, in small systems, accurate gates can be obtained at driving times competitive with the time taken by conventional sequences of two-qubit gates.

Moreover, we introduced the eigengate which maps the computational basis to the eigenbasis of some operator. This proves useful by making the resonant transitions between eigenstates relevant to quantum information processing, and as a Hahn echo which undoes dynamical phases accumulated due to a background Hamiltonian. We believe both the eigengate and our perspective on resonant driving may be of independent interest for other applications. We identified Polychronakos' model to feature an analytical eigengate as well as all the other requirements of our protocol, and showed numerically that the protocol does indeed work for small systems of size $N=4$ and $N=6$.

\section*{Acknowledgements}
We would like to thank Holger Frahm for pointing us to Polychronakos' model and Ref. \cite{Frahm1993}. This research was sponsored by the QM\&QI grant of the University of Amsterdam, supporting QuSoft.

\bibliographystyle{mybst}

\bibliography{Polychronakos}

\newpage
\appendix

\section{An intuitive picture of the halfway-inversion}
\label{app:hwp-intuition}

\begin{figure}[hbtp]
\begin{center}
\def\svgwidth{.45\linewidth}
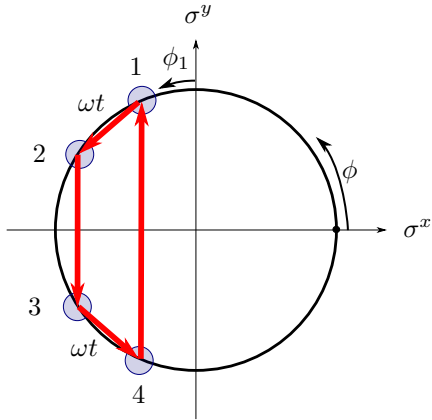
\caption{The relative phase $\phi$ between the two resonant states under our halfway-inversion protocol, depicted on the unit circle. Intuitively, the circle corresponds to the equatorial plane of the Bloch sphere, where any state on the sphere is first projected to the $\sigma^x,\sigma^y$ plane, and then projected to the nearest point on the sphere.  
An initial state which starts at the $+\sigma^z$ axis is first rotated towards the $+\sigma^y$ axis by resonant driving, whilst rotating around the Bloch sphere at frequency $\omega$. Thanks to the halfway inversion, the rotation by $\omega t$ is completely negated. }
\label{fig:hwp_driving_phases}
\end{center}
\end{figure}

In the main text, we claim that upon performing driving with halfway inversion (Eq. \ref{eqn:hwp-steps}), a perfect transition occurs in the case that $\delta = 0$. In this section, we indicate what happens to such a resonant pair of states, as indicated in Fig. \ref{fig:hwp_driving_phases}.

For two resonant states $\ket{t_1}$ and $\ket{t_2}$, we are interested in the relative phase on the Bloch sphere, $\phi$, defined as the relative phase in $\ket{t_1} + e^{i \phi} \ket{t_2}$. We follow what happens to an initial state $\ket{t_1}$, situated at the top of the Bloch sphere, at each of the protocol's steps. Note that a downside of this approach is that the relative phase $i \exp(\pm i \phi)$ given to the transitioning states (relative to the spectators) cannot be deduced. 
\begin{enumerate}
\item The first driving step, with phase $\phi_1$, drives the state towards the equator, at phase $\phi = \phi_1 + \frac{\pi}{2}$. For example, if $\phi=0$, a rotation around the $\sigma^x$ axis is performed, causing the state to end up at the $+\sigma^y$ axis. 
\item This, however, holds in the lab frame. Moving back to the rotating frame incurs a phase $\omega t$. Now, $\phi = \phi_1 + \frac{\pi}{2} + \omega t$. 
\item The halfway inversion's $\sigma^x$ effectively puts a minus sign in front of an equatorial state's phase. At this point, $\phi = -\phi_1 - \frac{\pi}{2} - \omega t$.
\item In the driving step that follows, the driving field's phase has been adjusted to $\phi_2 = -\phi_1 - \omega t$ in order to remain orthogonal to the state that should be inverted. In the rotating frame, no additional phase is accumulated, but the lab frame sees an increase of $\omega t$. At this point, $\phi = -\phi_1 - \frac{\pi}{2}$. 
\item The final $\sigma^x$ negates the phase again, such that the initial state $\ket{t_1}$ arrives at the bottom of the Bloch sphere ($\ket{t_2}$) at phase $\phi = \phi_1 + \frac{\pi}{2}$. 
\end{enumerate}

\end{document}